%% file: 2d-mtd_diffusion.tex
\documentclass[conference]{IEEEtran}
\IEEEoverridecommandlockouts

\usepackage{cite}
\usepackage{amsmath,amssymb,amsfonts}
\usepackage{graphicx}
\usepackage{textcomp}
\usepackage{xcolor}

\usepackage{algpseudocode}
\usepackage[linesnumbered,ruled,vlined]{algorithm2e}
\usepackage{subcaption}
\usepackage{graphicx}
\usepackage{hyperref}
\usepackage{float}
\usepackage{subcaption}
\usepackage{comment}

\DeclareMathOperator*{\argmax}{argmax}
\DeclareMathOperator*{\tr}{tr}

\def\BibTeX{{\rm B\kern-.05em{\sc i\kern-.025em b}\kern-.08em
    T\kern-.1667em\lower.7ex\hbox{E}\kern-.125emX}}

\begin{document}

\title{Score-based diffusion priors for multi-target detection}

\author{\IEEEauthorblockN{Alon Zabatani}
\IEEEauthorblockA{School of Electrical Engineering\\
Tel Aviv University\\
alonzabatani@mail.tau.ac.il}
\and
\IEEEauthorblockN{Shay Kreymer}
\IEEEauthorblockA{School of Electrical Engineering\\
Tel Aviv University\\
shaykreymer@mail.tau.ac.il}
\and
\IEEEauthorblockN{Tamir Bendory}
\IEEEauthorblockA{School of Electrical Engineering\\
Tel Aviv University\\
bendory@tauex.tau.ac.il}
}

\maketitle

\begin{abstract}
Multi-target detection (MTD) is the problem of estimating an image from a large, noisy measurement that contains randomly translated and rotated copies of the image.
Motivated by the single-particle cryo-electron microscopy technology, we design data-driven diffusion priors for the MTD problem, derived from score-based stochastic differential equations models.
We then integrate the prior into the approximate expectation-maximization algorithm.
In particular, our method alternates between an expectation step that approximates the expected log-likelihood and a maximization step that balances the approximated log-likelihood with the learned log-prior.
We show on two datasets that adding the data-driven prior substantially reduces the estimation error, in particular in high noise regimes. 
\end{abstract}

\begin{IEEEkeywords}
Diffusion models, multi-target detection, cryo-EM, score-SDE, expectation-maximization
\end{IEEEkeywords}

\section{Introduction}
\label{sec:introduction}
We study the two-dimensional multi-target detection (MTD) problem: estimating an image $f:\mathbb{R}^2 \rightarrow \mathbb{R}$ from a  single, noisy measurement that contains multiple copies of the image, each randomly rotated and translated~\cite{bendory2021multi,marshall2020image}. 
In particular, we model the measurement $M \in \mathbb{R}^{N \times N}$ as
\begin{equation}
\label{eq:model}
M[\vec{\ell}] = \sum_{i=1}^{p} F_{\phi_i}[\vec{\ell} - \vec{\ell}_i] + \varepsilon[\vec{\ell}],
\end{equation}
where \mbox{$F_{\phi_i} [\vec{\ell}] = f_{\phi_i} (\vec{\ell} / n)$} is a discrete copy of~$f$, rotated by angle~$\phi_i$ about the origin, $n$ is the known radius of the image in pixels, $\{\phi_i\}_{i=1}^{p}$ and $\{\vec{\ell}_i\}_{i=1}^{p}$ are the unknown rotations and translations, respectively, and $\varepsilon[\vec{\ell}]$ is i.i.d.\ Gaussian noise with zero mean and variance~$\sigma^2$.

\begin{figure}[!tb]
	\begin{subfigure}[ht]{0.30\columnwidth}
		\centering
		\includegraphics[width=\columnwidth]{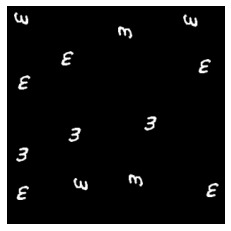}
		\caption{No noise}
	\end{subfigure}
	\hfill
	\begin{subfigure}[ht]{0.30\columnwidth}
		\centering
		\includegraphics[width=\columnwidth]{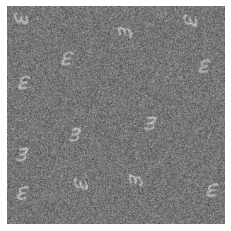}
		\caption{$\text{SNR} = 1$}
	\end{subfigure}
	\hfill
	\begin{subfigure}[ht]{0.30\columnwidth}
		\centering
		\includegraphics[width=\columnwidth]{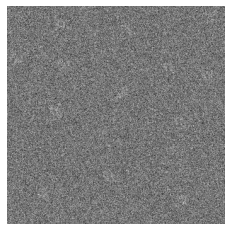}
		\caption{$\text{SNR} = 0.1$}
	\end{subfigure}
	\caption{Three measurements at different SNRs: (a)~no noise; (b)~\mbox{$\text{SNR} = 1$}; (c)~\mbox{$\text{SNR} = 0.1$}. Each measurement contains multiple rotated versions of the digit three, taken from the MNIST dataset. We are primarily interested in low SNR regimes {(e.g., panel~(c))}, in which it is extremely challenging to estimate the locations and rotations associated with the image occurrences.}
\label{fig:Micrographs_noise}
\end{figure}

The goal is to estimate the image $f$ from measurement~$M$, while
the rotations, translations, and the number of occurrences of~$f$ in~$M$, denoted by~$p$, are unknown. Importantly, since the rotations are unknown, it is possible to reconstruct the target image only up to a rotation. 
For simplicity,  we assume 
that the translations are not densely packed, namely, 
\begin{equation}
\label{eq:sep}
|\vec{\ell}_{i} - \vec{\ell}_{j}| > 4n, \quad \text{ for all } i \ne j.
\end{equation}
This assumption was also taken by previous works~\cite{marshall2020image, bendory2021multi} for simplification. The algorithm presented in this paper can be readily extended to handle arbitrarily spread signal occurrences, which violate~\eqref{eq:sep}, along with the guidelines that were proposed in~\cite{lan2020multi,kreymer2021two}.
Figure \ref{fig:Micrographs_noise} illustrates measurements with the digit three, taken from the MNIST dataset~\cite{lecun1998mnist}, at different signal-to-noise ratios (SNRs). We define $\text{SNR} \triangleq \frac{||F||^2_{\text{F}}}{A \sigma^2}$, where $A$ is the area in pixels of $F$.

The MTD model is mainly studied as a mathematical abstraction of the single-particle cryo-electron microscopy (cryo-EM) technology for determining macromolecular structures~\cite{henderson1995potential, nogales2016development, bai2015cryo}. In a cryo-EM experiment, individual copies of a target biomolecule, rotated by unknown 3-D rotations, are scattered across unknown 2-D locations within a thin layer of vitreous ice, from which 2-D tomographic projection images are produced by an electron microscope~\cite{frank2006three}.
Maintaining a low electron dose is imperative to prevent irreversible structural damage, leading to significantly noisy projection images. 

The conventional cryo-EM data processing pipeline \cite{bendory2020single, singer2020computational, scheres2012relion, punjani2017cryosparc} involves the initial detection and extraction of the 2-D projection images from the measurement, followed by their rotational and translational alignment to reconstruct the 3-D molecular structure. However, this approach fails for small molecules, which are too difficult to detect and align~\cite{henderson1995potential, aguerrebere2016fundamental,bendory2018toward,bendory2020single}. Consequently, the MTD model---that can be thought of as a simplification of the cryo-EM model, where 2-D rotations replace the 3-D rotations, and the tomographic projection operator is neglected---was devised in~\cite{bendory2018toward,bendory2019multi}, to study the recovery of small molecular structures, without intermediate detection, below the current detection limit~\cite{d2021current}.

Previous works proposed an expectation-maximization (EM) algorithm to solve the MTD problem ~\cite{lan2020multi, kreymer2022approximate}, outperforming an autocorrelation analysis alternative~\cite{marshall2020image, bendory2021multi, kreymer2021two}. EM is an iterative algorithm to find maximum likelihood (or maximum a posteriori) estimators, where the underlying model depends on unobserved nuisance parameters (in our case, the rotations and translations associated with the $p$ image occurrences). EM alternates between an expectation step (E-step), which calculates the expected log-likelihood evaluated using the current estimate of the parameters, and a maximization step (M-step), which finds the parameters maximizing the function computed in the E-step~\cite{dempster1977maximum}.
While EM cannot be applied directly to the MTD problem, due to a large number of possible translations that should be taken into account, the authors of \cite{lan2020multi} suggested an approximation scheme that dramatically reduces the computational burden, while providing high estimation quality; see Section~\ref{sec:map-em} for details.

Motivated by recent attempts to utilize prior information to constitute molecular structures in cryo-EM (see for example~\cite{kimanius2021exploiting, kimanius2023data}), we suggest incorporating score-based diffusion priors into the approximate EM framework. Diffusion models \cite{sohl2015deep, ho2020denoising, song2019generative, song2020score} are a class of probabilistic generative models that progressively perturb data by injecting noise, then learn to reverse this process, thus capturing the implicit prior of the underlying data distribution. Diffusion models have been successfully applied to a variety of tasks~\cite{dhariwal2021diffusion, amit2021segdiff, austin2021structured, alcaraz2022diffusion}, such as image synthesis, where they have broken the long-time dominance of generative adversarial networks (GANs)~\cite{goodfellow2014generative}. Score-based stochastic differential equations (score-SDE) \cite{song2020score} is a popular diffusion model, that makes use of the Stein score---the gradient of the log probability density\cite{liu2016kernelized, hyvarinen2005estimation}. The key idea behind score-SDE is to perturb the data with a sequence of intensifying Gaussian noise governed by a closed-form stochastic differential equation, and jointly estimate the score functions for all noisy data distributions by training a deep neural network,  conditioned on the noise levels.

The main contribution of this paper is the integration of score-based diffusion priors into the approximate EM framework, resulting in a flexible algorithm that balances between the log-likelihood term (the data) and the log-prior term using a regularized gradient ascent algorithm. We demonstrate on two datasets that we can successfully apply it to the two-dimensional MTD problem~\eqref{eq:model} and significantly improve image recovery quality, including in highly noisy regimes.

The rest of the paper is organized as follows.
Section~\ref{sec:score-sde} introduces the score-SDE method.
Section \ref{sec:map-em} presents the integration of score-SDE priors into the approximate EM framework.
Section \ref{sec:results} provides numerical results on two datasets (Gaussian images and MNIST), and Section~\ref{sec:conclusion} concludes the paper and delineates future applications to cryo-EM.

\section{Score-based Diffusion Models}
\label{sec:score-sde}
Score-based diffusion models learn the implicit prior of the underlying data distribution by matching the gradient of the log probability density with respect to data point $x$, i.e., Stein score function~\cite{liu2016kernelized, hyvarinen2005estimation}
\begin{equation}
\label{eq:stein-score}
    s(x) \triangleq \nabla_x \log p(x).
\end{equation}
We approximate the score function using a neural network $s_\theta(x)$, where the network’s parameters $\theta$ are being learned by minimizing the following score matching loss:
\begin{equation}
\label{eq:score-matching}
\text{loss}(\theta) = \mathbb{E}_{x \sim p(x)}  || s_\theta(x) - \nabla_x \log p(x) ||^2_2.
\end{equation}
However, in practice, the true gradients of the log probability density functions are often unknown during training. In~\cite{hyvarinen2005estimation}, it was shown that the loss function~\eqref{eq:score-matching} is equivalent to
\begin{equation}
\label{eq:score-matching-trace}
\text{loss}(\theta) = \mathbb{E}_{x \sim p(x)} \left(\frac{1}{2}|| s_\theta(x) ||^2_2 + \tr\left(\nabla_x s_\theta(x)\right)\right),
\end{equation}
where $\text{tr}(\cdot)$ is the trace of the Jacobian matrix $\nabla_x s_\theta(x)$.
Unfortunately, learning the score function by minimizing the loss function~\eqref{eq:score-matching-trace} tends to fail  in regions where there are fewer training samples (due to the expectation over the data density);
this can be mitigated by perturbing the data with various noise levels, as shown in \cite{song2019generative}.
Specifically, score-SDE \cite{song2020score} perturbs the data with a sequence of intensifying Gaussian noise, governed by a closed-form stochastic differential equation, and jointly estimates the score functions for all noisy data distributions. The SDE is defined as follows
\begin{equation}
\label{eq:sde}
dx = - \frac{\beta (t)}{2}x dt + \sqrt{\beta (t)}dw,
\end{equation}
where $\beta (t): \mathbb{R} \rightarrow \mathbb{R} > 0$ is the noise schedule of the process, typically taken to be a monotonically increasing linear function of $t$, and $w$ is the standard Wiener process. The data distribution is defined when $t = 0$, i.e., $x(0) \sim p_{data}$, and a simple, tractable distribution (e.g., Gaussian) is assumed when $t=T$, i.e., $x(T) \sim \mathcal{N}(0, I)$.

With the evolution of the distribution along time $t$, the neural network accepts both data point $x$ and time $t$ as inputs and produces an estimate of the score function, i.e., $s_\theta(x,t)$. In this case, the denoising score-matching loss function can be written as
\begin{multline}
\label{eq:denoising-score-matching}
    \text{loss}(\theta) = \mathbb{E}_{t \sim U[0, T]} \mathbb{E}_{p(x(0))} \mathbb{E}_{p(x(t) | x(0))} \\ \Big( \lambda (t) || s_\theta(x(t), t) - \nabla_{x(t)} \log p(x(t)|x(0)) ||_2^2 \Big),
\end{multline}
where $\lambda(t)$ is a positive weighting function and $p(x(t)|x(0))$ can be analytically calculated using~\eqref{eq:sde} to be:
\begin{equation}
p(x(t)|x(0)) \sim \mathcal{N}(\mu (t), \sigma^2(t) I),
\end{equation}
where
\begin{align*}
\mu(t) = x(0) \exp \left(-\frac{1}{2}\int_0^t \beta(s) ds\right),
\\
\sigma^2(t) = 1 - \exp \left(-\int_0^t \beta(s) ds \right).
\end{align*}
The expectations are approximated by averaging over the data points and the integrals can be solved analytically for the specified functions (full implementation details can be found in Table~\ref{tbl:parameters}).

\section{Approximate EM with score-based prior}
\label{sec:map-em}
The EM algorithm finds a (local) maximum of the posterior distribution (the MAP estimator) by iteratively applying an expectation step (E-step), followed by a maximization step (M-step). The E-step calculates the expectation of the log-likelihood with respect to all admissible configurations of nuisance parameters (translations and rotations in our case). The M-step then finds the parameters maximizing the function found in the E-step, possibly with the addition of a prior term \cite{dempster1977maximum}. 

Direct application of the EM algorithm to the MTD problem is computationally intractable, due to the large number of possible translations that grows quickly with the measurement size. Specifically, under the separation constraint \eqref{eq:sep}, the runtime of a single iteration grows as $\mathcal{O}(N^4)$, where $N^2$ is the size of the measurement M \eqref{eq:model}, assuming the measurement is much larger than the target image.
Approximate EM that has been introduced in~\cite{lan2020multi} runs in $\mathcal{O}(N^2)$ per iteration, by maximizing an approximation of the log-likelihood function. 
The approximate EM algorithm has been applied to the two-dimensional MTD problem in~\cite{kreymer2022approximate} and to the cryo-EM problem in~\cite{kreymer2023stochastic}. However, in both cases, the prior on the signal has been ignored, which is equivalent to considering an uninformative prior.
The main goal of this work is to introduce a data-driven prior into the approximate EM algorithm and to pave the way to learn and implement such prior for cryo-EM datasets.

\begin{figure}[tbp]
    \centering
    \captionsetup{justification=centering}
    \subfloat[A target image $F$ of size~$L = 28$ pixels.]{
        \includegraphics[width=0.28\columnwidth, keepaspectratio]{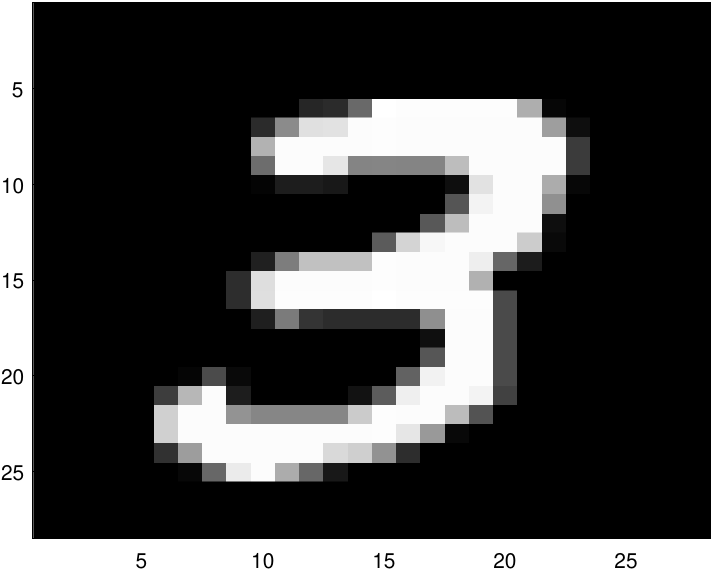}
    \label{fig:F}}
    \hfill
    \subfloat[The padded image $ZF$ of size $2L = 56$ pixels.]{
        \includegraphics[width=0.28\columnwidth, keepaspectratio]{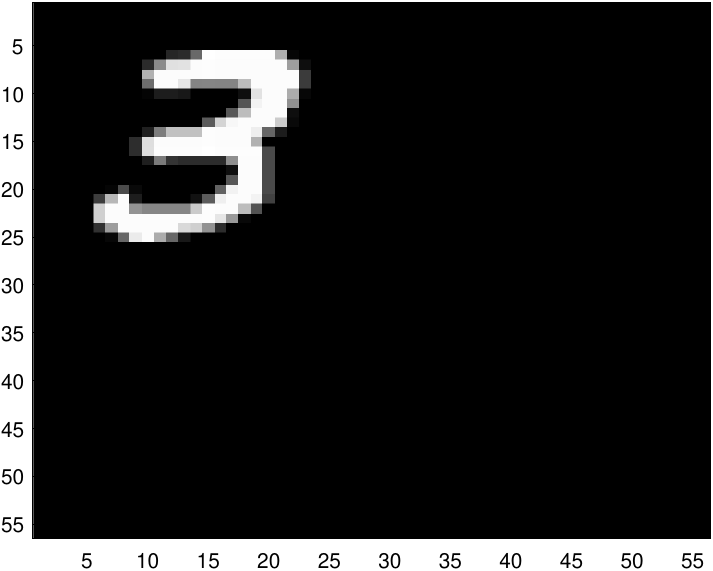}
        \label{fig:ZF}}
    \hfill
    \subfloat[The padded and shifted image $T_{\vec{\ell_m}} Z F$,  shifted by $\vec{\ell_m}= (7, 10)$ pixels.]{
        \includegraphics[width=0.28\columnwidth, keepaspectratio]{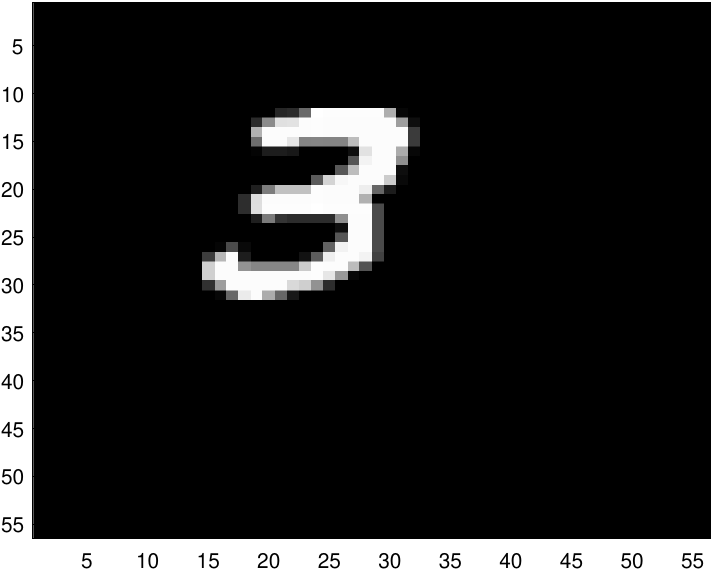}
        \label{fig:TZF}}
    
    \medskip
    
    \subfloat[The cropped image $C T_{\vec{\ell_m}} Z F$ of size $L = 28$ pixels.]{
    \includegraphics[width=0.28\columnwidth, keepaspectratio]{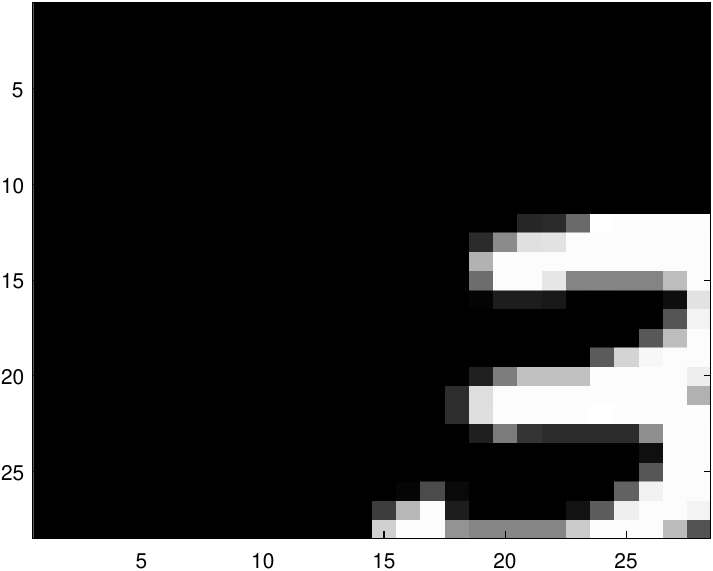}
    \label{fig:CTZF}}
    \quad
    \subfloat[The noisy patch $M_m$ with $\text{SNR} = 4$.]{
    \includegraphics[width=0.28\columnwidth, keepaspectratio]{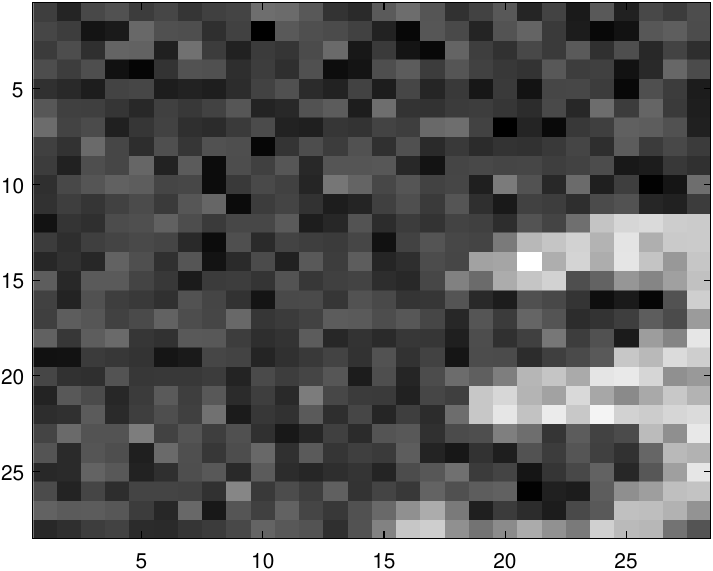}
    \label{fig:CTZF_E}}
    
    \caption{A step-by-step illustration of the patch generation model described in~\eqref{eq:patch}.}
    \label{fig:patch}
\end{figure}

Approximate EM begins with partitioning the measurement M into $D = (N/L)^2$ non-overlapping patches, each of size $L \times L$ (we assume for convenience that $N/L$ is an integer). The separation condition implies that each patch $M_m, m=0\ldots,D-1$, can contain either no target image, a full  target image, or a part of the target image; overall there are~$(2L)^2$  possibilities, excluding rotations. In particular, each patch can be modeled by
\begin{equation}
\label{eq:patch}
M_m = C T_{\vec{\ell}_m} Z F_{\phi_m} + \varepsilon_m, \quad \varepsilon_m \sim \mathcal{N}(0, \sigma^2 I),
\end{equation}
where~$F$ is a square image of size $L \times L$, representing a discrete copy of target image $f$, the operator~$Z$ \mbox{zero-pads}~$L$ entries to the right and to the bottom of a rotated copy of~$F$, and~$T_{\vec{\ell}_m}$ circularly shifts the \mbox{zero-padded} image by~\mbox{$\vec{\ell}_m = ({\ell_m}_x, {\ell_m}_y)\in \mathbb{L} \triangleq \{0, 1, \ldots, 2L-1\}^2$} positions, that is,
\begin{multline}
(T_{\vec{\ell}_m} Z F_{\phi_m} )\left[i, j\right] = \\(Z F_{\phi_m}) \left[(i + {\ell_m}_x) \bmod 2L, (j + {\ell_m}_y) \bmod  2L\right].
\end{multline}
The operator~$C$ then crops the image to size $L \times L$, i.e., \mbox{$CT_{\vec{\ell}_m} Z F_{\phi_m} = T_{\vec{\ell}_m} Z F_{\phi_m}[0:L-1, 0:L-1]$}, and the result is further corrupted by additive white Gaussian noise. Figure~\ref{fig:patch} provides an illustration of the model.
In addition, since the EM algorithm assigns probabilities to rotations (in the E-step), we uniformly discretize the search space of rotations:
\begin{equation}
\label{eq:Phi_set}
\phi_m \in \Phi \triangleq \left\{k \frac{2\pi}{K}\right\}, \quad k=0,\ldots,K-1,
\end{equation}
where~$K$ is a hyper-parameter for the algorithm.

The patches $\{M_m\}_{m=0}^{D-1}$ are our observations, whereas the rotations $\{\vec{\ell}_m\}_{m=0}^{D-1}$ and translations $\{\phi_m\}_{m=0}^{D-1}$ are the unobserved nuisance parameters. Our goal is thus to find the pixel intensity values of $F$, and we denote by $F^{(\tau)}$ the estimated image at iteration $\tau$ of the EM procedure.
In the E-step, we calculate the expectation of log-likelihood function of the model, with respect to the conditional distribution of the nuisance parameters given the patches and the current estimate $F^{(\tau)}$:

\begin{equation}
Q(F | F^{(\tau)}) \triangleq \mathbb{E}\left[\log \mathfrak{L} \big| M_0, \ldots, M_{D-1}; F^{(\tau)} \right],
\end{equation}
where $\mathfrak{L}$ is the likelihood function, defined as
\begin{multline}
\label{eq:approx_likelihood}
\mathfrak{L} \triangleq p(M_0, \ldots, M_{D-1}, \vec{\ell}_0, \ldots, \vec{\ell}_{D-1}, \phi_0, \ldots, \phi_{D-1};F) \\\approx \prod_{m = 0}^{D - 1} p(M_m, \vec{\ell}_m, \phi_m;F),
\end{multline}
where we neglect statistical dependencies between patches. 

We note that Bayes' rule implies
\begin{multline}
p(\vec{\ell}_m, \phi_m|M_m, F^{(\tau)}) \\= \frac{p(M_m|\vec{\ell}_m, \phi_m, F^{(\tau)}) p(\vec{\ell}_m, \phi_m|F^{(\tau)})}{\sum_{\vec{\ell'} \in \mathbb{L}} \sum_{\phi' \in \Phi} p(M_m|\vec{\ell'}, \phi', F^{(\tau)}) p(\vec{\ell'}, \phi'|F^{(\tau)})},
\end{multline}
where from \eqref{eq:patch} we can write
\begin{equation}
\label{eq:likelihood_patch}
p(M_m|\vec{\ell}_m, \phi_m, F) \propto \exp \left(- \frac{\|M_m - C T_{\vec{\ell_m}} Z F_{\phi_m}\|_\text{F}^2}{2 \sigma^2} \right),
\end{equation}
with the normalization~\mbox{$\sum_{\vec{\ell} \in \mathbb{L}} \sum_{\phi \in \Phi} p(M_m|\vec{\ell}_m, \phi_m, F) = 1$}.
We assume that in each patch~$p(\vec{\ell}_m,\phi_m|F)=p(\vec{\ell}_m)p(\phi_m)=\frac{1}{(2L)^2} \cdot \frac{1}{K}$, namely,~$\vec{\ell}_m$ and~$\phi_m$ are independent of~$F$ and of each other, and are uniformly distributed. Finally, utilizing the approximation in~\eqref{eq:approx_likelihood}, we can write the expected log-likelihood function, up to a constant, as:
\begin{multline}
\label{eq:Q_function}
Q(F|F^{(\tau)}) = \sum_{m = 0}^{D - 1} \sum_{\vec{\ell} \in \mathbb{L}} \sum_{\phi \in \Phi} p(\vec{\ell}_m, \phi_m|M_m, F^{(\tau)}) \\ \times \log p(M_m|\vec{\ell}_m, \phi_m, F).
\end{multline}

The~\mbox{M-step} updates the current image estimate~$F^{(\tau)}$ by maximizing the Q-function \eqref{eq:Q_function}, with the addition of a regularized prior term:
\begin{equation}
	\label{eq:M_step}
	F^{(\tau+1)} = \argmax_{F} \left( Q(F|F^{(\tau)}) + \gamma^{(\tau)} \log p(F) \right),
\end{equation}
where $\gamma^{(\tau)}$ is a weighting function, applied to the prior term at iteration $\tau$.
In our numerical experiments, the maximization process is being conducted using a fixed number of gradient ascent steps with a fixed learning rate $\mu$ and the initial guess of $F^{(\tau)}$.
The gradient of the Q-function can be analytically computed, whereas the gradient of the prior term is approximated by our diffusion model $s_\theta(\cdot)$.
Thus, a single gradient ascent step can be written as:
\begin{equation}
\begin{split}
F^{(\tau+1)} &= F^{(\tau)} + \mu \nabla_{F} \left( Q(F|F^{(\tau)}) + \gamma^{(\tau)} \log p(F) \right)\Big\rvert_{F = F^{(\tau)}} \\
&= F^{(\tau)} + \mu \left( \nabla_{F} Q(F|F^{(\tau)})\rvert_{F = F^{(\tau)}} + \gamma^{(\tau)} s_\theta(F^{(\tau)}) \right).
\label{eq:gradient-ascent}
\end{split}
\end{equation}
The algorithm is summarized in Algorithm \ref{alg:approx_EM}.

\begin{algorithm}[!tb]
  \caption{Approximate EM with score-based prior}\label{alg:approx_EM}
\KwIn{measurement~$M$; noise variance~$\sigma^2$; initial guess~$F^{(0)}$; rotation range~$K$~\eqref{eq:Phi_set}; max EM iterations $\mathcal{T}$; stopping parameter~$\epsilon$; learning rate $\mu$; regularization weights $\gamma$; prior model $s_\theta(\cdot)$; gradient ascent steps $N$\\}
\KwOut{an estimate of~$F$}
\BlankLine
  \For{$\tau \gets 0$ to $\mathcal{T}$}{
    calculate $\nabla_F Q(F|F^{(\tau)})$ according to~\eqref{eq:Q_function}\\
    set $F_0 \leftarrow F^{(\tau)}$\\
    \For{$i \gets 0$ to $N$}{
    set $F_{i+1} \leftarrow F_i + \mu \left( \nabla_{F} Q(F_i|F^{(\tau)}) + \gamma^{(\tau)} s_\theta(F_i) \right)$\\
    }
    set $F^{(\tau+1)} \leftarrow F_N$\\
    \If{$||F^{(\tau+1)} - F^{(\tau)}||_{\text{F}} < \epsilon$} {
    break}
  }
  return $F^{(\tau)}$
\end{algorithm}

\section{Experimental Results}
\label{sec:results}
In this section, we implement our proposed algorithm on two different datasets: (A) Gaussian $5 \times 5$ images ($L=5$) and (B) MNIST $28 \times 28$ images ($L=28$).
As our baseline, we consider the plain approximate EM framework (i.e., without diffusion priors) that has been proposed in~\cite{kreymer2022approximate}. We examine both algorithms on a range of SNR values scaled for each dataset; for example, the sparsity of the MNIST images requires lower SNR compared to Gaussian images so that the target images are not easily detected (as in panel (c) of Figure \ref{fig:Micrographs_noise}). Specifically, for the Gaussian case we test $\text{SNRs} = \{1, 5, 10\}$, and for the MNIST case we test $\text{SNRs} = \{0.1, 0.5, 1\}$.
We generate the measurements according to the MTD model~\eqref{eq:model}, with $N=11L$ and density $\frac{pL^2}{N^2} = 0.1$. The rotations are uniformly sampled from the set $\Phi \triangleq \{0, \frac{\pi}{2}, \pi, \frac{3\pi}{2}\}$, so that $K=4$ \eqref{eq:Phi_set}. Finally, for each dataset we examine 10 target images (that are hidden from the prior models during training) and report the mean estimation error, defined by $E = \min_{\phi \in \Phi} \frac{||F^* - F_{\phi} ||_2}{||F^*||_2}$, where $F^*$ denotes the ground truth image and $F_\phi$ denotes the estimated image, rotated by~$\phi$. The error is minimized over the set of rotations because of the intrinsic rotational symmetry of the problem. Additional implementation details are provided in Table \ref{tbl:parameters}. The code to reproduce the experiments is available at \url{https://github.com/zabatani/2d-mtd_diffusion}.

\begin{figure}[tbp]
	\begin{subfigure}[ht]{\columnwidth}
		\centering
		\includegraphics[width=\columnwidth, keepaspectratio]{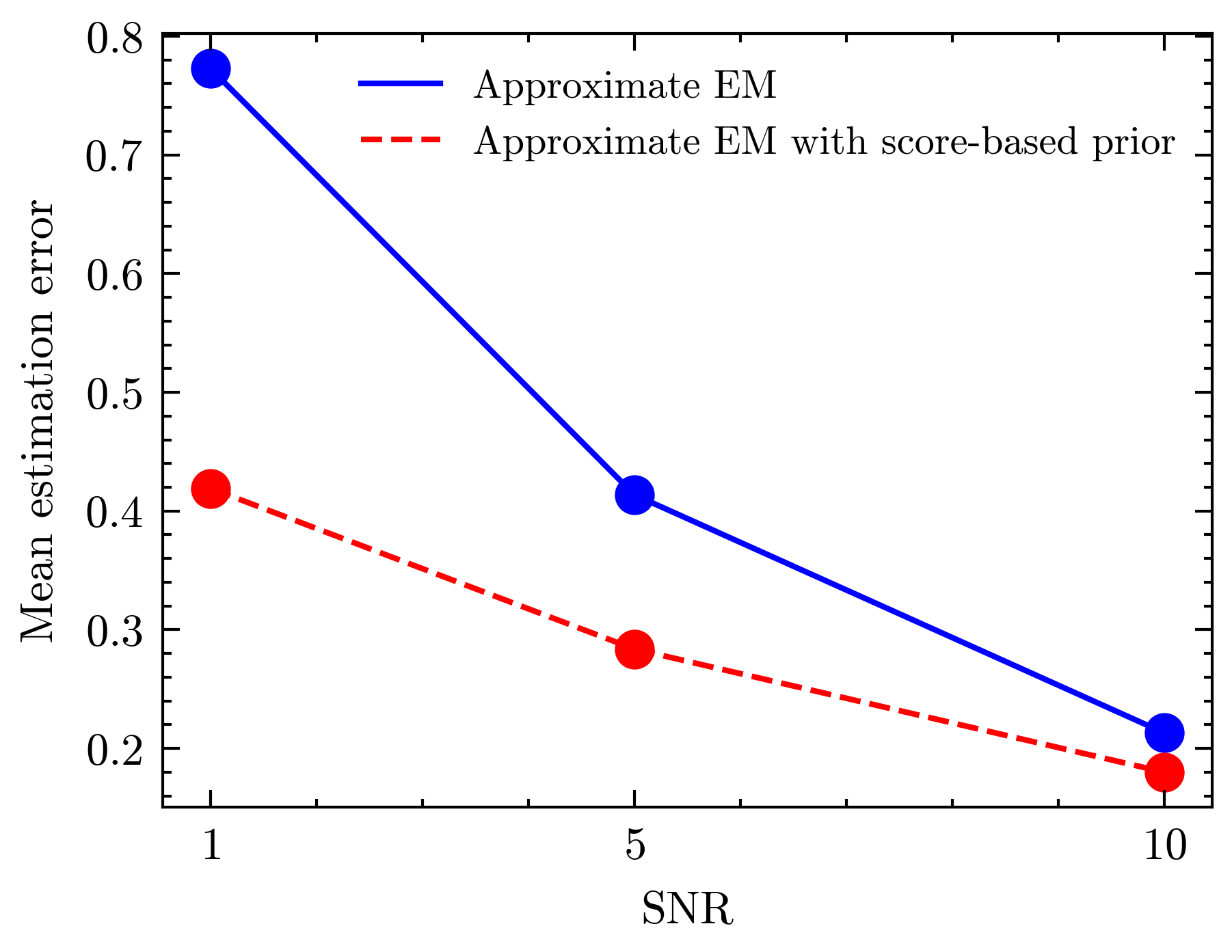}
	\end{subfigure}
	\caption{The mean estimation error of recovering target images~$F$ drawn from a Gaussian distribution, as a function of SNR, using approximate EM with and without score-based prior. Evidently, the score-based prior addition is beneficial in all SNRs, especially in low SNR regions.}
\label{fig:gaussian_snr}
\end{figure}

\subsection{Gaussian Images}
\label{res:gaussian}
The score network for this dataset consists of three fully-connected layers with Softplus activation functions. The network has been trained on 10,000 samples drawn from a normal distribution with covariance identity and mean $\mu$, where each entry of $\mu$ has been drawn i.i.d. from a uniform distribution over $\{0,1,2,3,4\}$. The target images were drawn from the same distribution. 
Training took approximately 10 minutes on an M1 Macbook Air. 
The results are illustrated in Figure~\ref{fig:gaussian_snr}.  Evidently, our algorithm is superior in all SNRs and exhibits a significant improvement in recovery accuracy, especially in low SNR levels.

\begin{figure}[tbph]
	\begin{subfigure}[ht]{\columnwidth}
		\centering
		\includegraphics[width=\columnwidth, keepaspectratio]{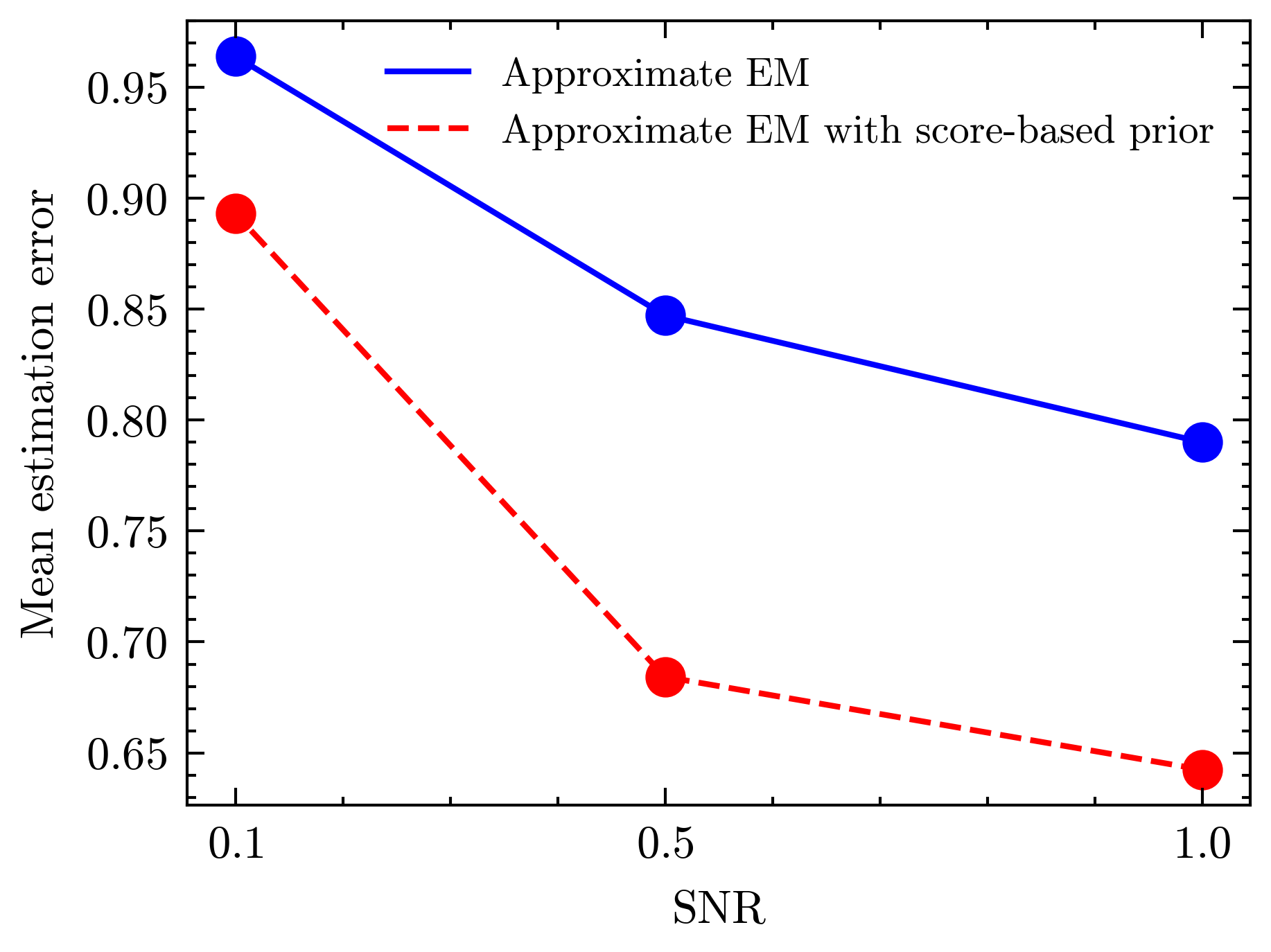}
	\end{subfigure}
	\caption{The mean estimation error of recovering target images $F$ drawn from the MNIST dataset, as a function of SNR, using approximate EM with and without score-based prior. Evidently, the score-based prior addition is noticeably beneficial in all SNRs. 
 }
\label{fig:mnist_snr}
\end{figure}

\subsection{MNIST}
\begin{figure*}[t]
\begin{subfigure}{\textwidth}
    \centering
    \includegraphics[width=\textwidth]{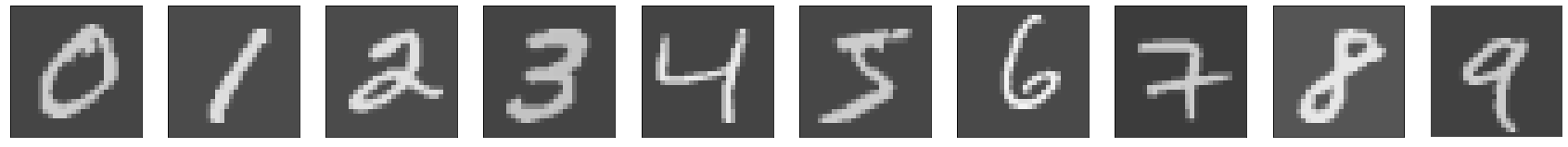}
    \caption{Target image}
\end{subfigure}
\begin{subfigure}{\textwidth}
    \centering
    \includegraphics[width=\textwidth]{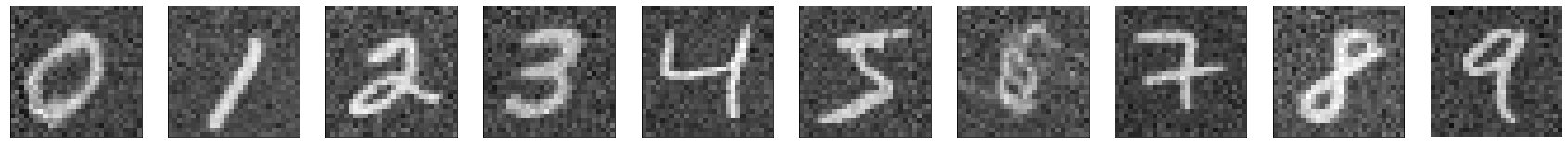}
    \caption{Approximate EM without prior}
\end{subfigure}
\begin{subfigure}{\textwidth}
    \centering
    \includegraphics[width=0.98\textwidth]{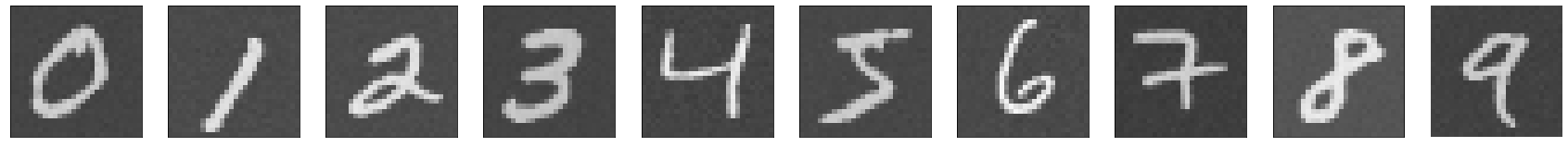}
    \caption{Approximate EM with score-based prior}
\end{subfigure}
\caption{Recovery of all ten digits from the MNIST dataset of the MTD model with $\text{SNR}=1$, using the approximate EM algorithm with and without score-based prior.  The recovery quality boost achieved by our algorithm (panel (c)) is clearly visible. Besides the background denoising and digit sharpening, it also successfully recovers the target image in cases where the competitor deteriorates (e.g., digit six).}
\label{fig:mnist_digit}
\end{figure*}

The score network for this dataset consists of a U-Net architecture \cite{ronneberger2015u}, with ten convolutional layers and LogSigmoid activation functions. The network has been trained on the entire MNIST dataset (60,000 samples), excluding the 10 target images that have been tested. 
The training took approximately 10 hours on a T4 GPU machine. In this case, we have noticed that proper weighting of the prior term during gradient ascent plays a key role in our algorithm's success and that a monotonically increasing linear function of $\tau$ (the iteration of the EM procedure) works best; see Table \ref{tbl:parameters} for details. The results are illustrated in Figure \ref{fig:mnist_snr}. Our algorithm is superior in all SNR levels. In addition, we show reconstructions of all digits for a fixed $\text{SNR} = 1$, using both algorithms, in Figure~\ref{fig:mnist_digit}. The improvement in recovery quality is visible for all digits.

\section{Conclusion}
\label{sec:conclusion}

In this paper, we proposed to integrate score-based diffusion priors into the approximate EM framework in order to estimate images in the two-dimensional MTD problem. Our method shows a superior image recovery quality compared to~\cite{kreymer2022approximate}, which considers an uninformative prior, in various SNRs and on two datasets: a simple Gaussian case, and the more intricate data distribution of MNIST.

This paper is part of an ongoing effort to recover small molecular structures using cryo-EM. The MTD model~\eqref{eq:model} is a simplified model of the cryo-EM model that includes 3-D rotations (rather than 2-D rotations in~\eqref{eq:model}) and tomographic projections~\cite{bendory2020single,singer2020computational}. 
The ultimate goal of this research is to learn a prior of protein structures that were resolved using electron microscopes, and then incorporate it into a computational method for recovering small molecular structures~\cite{bendory2018toward,kreymer2023stochastic}. To learn the priors, we intend to devise 3-D steerable CNN architectures~\cite{weiler20183d} and train them in a denoising score-matching fashion \eqref{eq:denoising-score-matching}. This would require collecting enough data to train such neural networks. To this end, we plan to use databases of resolved protein structures, such as the electron microscopy data bank (EMDB)~\cite{lawson2016emdatabank}, as well to augment data generated by computational tools, such as AlphaFold~\cite{jumper2021highly,terwilliger2023alphafold}. Moreover, we intend to explore additional ways to incorporate the score function into the EM framework that might take advantage of the reverse-time SDE~\eqref{eq:sde} (commonly used for sample generation), for example, by a clever integration between the SDE time evolution and the EM iteration progression. Finally, we intend to integrate our prior into additional algorithms for MTD, such as autocorrelation analysis~\cite{marshall2020image, bendory2021multi,bendory2018toward,kam1980reconstruction}.

\section*{Acknowledgment} 
The research is supported in part by the BSF grant no. 2020159, 
the NSF-BSF grant no. 2019752, and the ISF grant no. 1924/21. 

\begin{table*}
\centering
\caption{The parameters used for the numerical experiments.}
\begin{tabular}{||c c c c||}
 \hline
 Parameter & Description & Gaussian & MNIST \\
 \hline\hline
 $\mathcal{T}$ & EM iterations & 100 & 100 \\
 \hline
  $N$ & Gradient ascent steps & 1 & 1 \\  
 \hline
 $\gamma^{(\tau)}$ & Regularization weighting function \eqref{eq:gradient-ascent} & 1 & $\tau/\mathcal{T}$ \\ 
 \hline
  $\mu$ & Learning rate \eqref{eq:gradient-ascent} & $1e-3$ & $1e-3$ \\ 
 \hline
 $\epsilon$ & Stopping parameter & $1e-5$ & $1e-5$ \\ 
 \hline
 $F^{(0)}$ & Initialization & $U[0, 1]$ & Average of ten digits \\
 \hline
$\text{loss}(\theta)$ & Prior loss function & \eqref{eq:score-matching-trace} & \eqref{eq:denoising-score-matching} \\
 \hline
  $\beta(t)$ & Noise schedule \eqref{eq:sde} & N/A & $0.1 + (20-0.1)t$ \\
 \hline
  $\lambda(t)$ & Weighting function \eqref{eq:denoising-score-matching} & N/A & $1-e^{\int_0^t \beta(s)ds}$ \\
 \hline
\end{tabular}
\label{tbl:parameters}
\end{table*}

\bibliographystyle{IEEEtran}
\input{references.bbl}

\end{document}